\documentclass[12pt,a4paper]{article}
\usepackage[final]{showkeys}
\usepackage{amsmath}
\usepackage{amssymb}
\usepackage{amsthm}

\renewcommand{\t}[1]{\texttt{#1}}

\title{ A Unified Picture of Decoherence Control}
\author{R.~Alicki~\footnote{Electronic address: 
\t{fizra@univ.gda.pl}} \\
{\normalsize Institute of Theoretical Physics and Astrophysics} \\
{\normalsize University of Gda\'nsk, Poland}} 
\begin{document}

\maketitle
\begin{abstract}
The different methods of reducing decoherence in quantum devices are discussed from the unified point of view
based on the energy conservation principle and the concept of forbidden transitions. Minimal decoherence model, "bang-bang" techniques, Zeno effect and decoherence-free subspaces and subsystems are studied as particular examples.
\end{abstract}

\section{Introduction}
The rapid development of quantum information theory, both in its theoretical and experimental aspects \cite{NC,A} has posed new 
and renew old questions and challenges in the theory of quantum open systems \cite{BF,J}. Fulfilling  extreme demands concerning the accuracy of controlled operations ({\em unitary gates}) performed on microscopic quantum devices is crucial for any attempts
to implement the ideas of quantum computation or generally quantum information processing at a reasonable scale.

One can find in the literature different ideas of various technical complexity which should be useful for reducing errors due to decoherence processes in quantum systems interacting with an environment. We shall concentrate our attention on four of them : minimal decoherence model \cite{RA,AHHH}, "bang-bang" techniques \cite{VL}, Zeno effect \cite{MS,FP} and decoherence-free subspaces and subsystems \cite{ZF,LW}. The whole field of active quantum error correction methods
\cite{K} is not discussed here. An interesting question remains, whether these methods can be explained in a similar fashion also.   

Mathematical relations between different methods of noise reduction in quantum systems have been already discussed in the literature \cite{ZLF}. The main goal of this paper is to show that those different techniques can be understood from a unified physical point
of view of two elementary principles - the energy conservation and  the existence of forbidden transitions. Moreover, for the quantitative approach only the elementary time-dependent perturbation theory ({\em Fermi Golden Rule}) is used.
One should mention, however, that for some cases more advanced and rigorous methods are known in the literature.

\subsection{Fermi Golden Rule}

Consider a quantum system with a Hamiltonian $H_0$ perturbed by a "small" term $V$ such that the total Hamiltonian reads
\begin{equation}
H= H_0 +  V\ .
\label{ham} 
\end{equation}  
Assuming the spectral resolution of $H_0$ in the form
\begin{equation}
H_0 |k> =  E_k |k>
\label{gen} 
\end{equation}
one can obtain the celebrated {\em Fermi Golden Rule}  for the {\em transition probability per unit time} from the state
$|m>$ to the state $|n>$ caused by the small perturbation: 
\begin{equation}
P_{nm} = \frac{2\pi}{\hbar}|<n|V|m>|^2\delta(E_n -E_m )\ .
\label{Fer}
\end{equation}
The formula (\ref{Fer}), as stated, is rather singular and needs some additional explanation. In particular, taking sums over certain sets of initial and final states is usually necessary. However, some general rules concerning the suppression of decoherence
can be extracted directly from (\ref{Fer}). To avoid decoherence for a set of initial states $\{|m>\}$ we must put the transition
probability from  $|m>$ to all other states  equal to zero. This is the case if the energy conservation $E_n =E_m $
cannot be fulfilled or the transitions from any initial $|m>$ to all other states are forbidden, i.e. $<n|V|m>= 0$.

It will be useful to notice that the transition probability $P_{nm}$ can be written as an integral 
\begin{equation}
P_{nm} = \frac{1}{\hbar^2}\int_{-\infty}^{\infty}V_{nm}(t) V_{mn}(0)\,dt
\label{Fer1}
\end{equation}
where $V_{nm}(t) = <n|V(t)|m>= \exp\{i(E_n -E_m)t/\hbar\}<n|V|m>$  and $V(t) = e^{iH_0t/\hbar}V e^{-iH_0t/\hbar}$. 
In each particular application of (\ref{Fer}) we have to average the transition probability over certain sets of initial and final quantum states and then (\ref{Fer1}) becomes an example of Kubo formula for transport coefficients expressed as integrals of {\em autocorrelation functions} \cite{KTH}. 
\section{Dynamical decoupling methods}
It is known from the classical theory of dynamical systems that a fast periodic perturbation can either enforce chaotic behavior in the otherwise regular system or in other cases stabilize the system evolution. The later phenomenon has been extended into quantum domain and proposed as a method for reducing decoherence effects in quantum information processing.  A rather opposite approach utilizes decrease of decoherence rate with an increasing time scale of quantum evolution. This type of behavior is characteristic for the systems linearly coupled to quantum fields (photons, phonons, etc.).
 
\subsection{Open system in weak coupling regime}

Consider an open system ${\cal S}$ weakly coupled to a large reservoir ${\cal R}$ . The total Hamiltonian is given by
\begin{equation}
H = H_S + H_R + H_{int}\ ,\  H_{int} = S\otimes R\ .
\label{open}
\end{equation}
The form $S\otimes R$ for the interaction Hamiltonian ($S=S^{\dagger}$ - system's operator, $R=R^{\dagger}$ -reservoir's
operator) is chosen for simplicity only. Most of the results can be easily generalized to the interactions of the form $\sum S_{\alpha}\otimes R_{\alpha}$. We assume a discrete spectrum for $H_S$ and a practically continuous one for $H_R$ 
\begin{equation}
H_S|k> = \epsilon_k |k>\ ,\  H_R |E,\gamma> = E |E,\gamma>\ .
\label{spop}
\end{equation}
with $\gamma$ describing additional degeneracy of energy levels.

We compute $P_{kl}$ - "the transition probability per unit time from the state $|l>$ to the state $|k>$ of the system {\cal S}" using the Fermi Golden Rule (\ref{Fer}) for the perturbation applied to the initial
state $|l>\otimes|E,\gamma>$ and to a final state $|k>\otimes|E',\gamma'>$. Averaging over the initial probability distribution for the reservoir states $\sigma (E,\gamma)$ and over all possible final states of ${\cal R}$ we obtain
\begin{equation}
P_{kl} = \frac{2\pi}{\hbar}|<k|S|l>|^2 
\nonumber
\end{equation}
\begin{equation}
\times\sum_{\gamma',\gamma}\int dE'\int dE\, \sigma(E,\gamma)|<E',\gamma'|R|E,\gamma>|^2\delta(\epsilon_l + E - \epsilon_k - E')\ .
\label{prob}
\end{equation}
The formula (\ref{prob}) will be used to discuss two oposite strategies of noise reduction: minimal decoherence model
and "bang-bang" techniques.

\subsection{Minimal decoherence model}
For many physical models the linear coupling of ${\cal S}$ to a quantum bosonic field, or, in other words to a system of quantum harmonic oscillators, is an important source of noise.
For these models the transition between two states $|l>$ and $|k>$ is accompanied by the emission or absorption
of a single boson (e.g. photon, phonon, etc.) of energy $\hbar\omega$. 
Therefore,
the allowed transitions for which $|<E,\gamma|R|E',\gamma'>|^2 > 0$ are only between the states of the quantum field
which differ by a single boson of frequency $\omega$, i.e. $|E-E'|= \hbar\omega$. The energy conservation leads to the condition
\begin{equation} 
\hbar\omega = |\epsilon_k -\epsilon_l|\ .
\label{fre}
\end{equation}
and hence, the transition probability
$P_{kl}$ is proportional to the density of the bosonic states $n(\omega)$ at $\omega = |\epsilon_k -\epsilon_l|/\hbar$.
Typically, such a density behaves like $\omega^r $, $r>0$, i.e. tends to zero for low frequencies. This suggests a {\em minimal decoherence strategy} which demands to use systems with almost degenerated energy levels and to apply slow gates producing sufficiently small energy level splitting - $\max |\epsilon_k -\epsilon_l|$.

There exist natural restrictions for application of minimal decoherence strategy. First of all, applying slow gates means using low frequency external (typically electromagnetic) fields for the system control. Low frequencies mean long waves
of the corresponding fields, which implies difficulties in individual control of localized constituents ("qubits") of the quantum device. Secondly, beyond the processes of emission or absorption of energy quanta, various scattering processes are always present. Their kinematics is completely different. For example, in the case of elastic scattering we can have
$|\epsilon_k -\epsilon_l|=0$, but the density of  outcoming scattered states, which replaces the density of bosons from the previous model, is different from zero.

\subsection{Bang-bang techniques}

These techniques utilize the typical behavior of the matrix elements\\  $<E,\gamma|R|E',\gamma'> $. Practically, for all known models of open systems\\ $<E,\gamma|R|E',\gamma'>\simeq 0$ if $E'-E >> E_{cut}$ for a certain cut-off energy $E_{cut}$.\\
Introducing a rapidly varying Hamiltonian $H_S(t)$ we can use a modified derivation of (\ref{prob}). In the interaction picture with respect to the dynamics governed by $H_S(t)+H_R$ we obtain the following time-dependent effective Hamiltonian
\begin{equation}
H_{int}(t) = S(t)\otimes R(t)
\label{hint}
\end{equation}
where $S(t), R(t)$ are the solutions of the Heisenberg equations
\begin{equation} 
\frac{d}{dt}S(t) = \frac{i}{\hbar}[H_S(t), S(t)]\ ,\ \frac{d}{dt}R(t) = \frac{i}{\hbar}[H_R, R(t)]\ .
\label{sint}
\end{equation}
We use a modification of the formula (\ref{Fer1}) which takes into account the time-dependence of $H_S(t)$.
This leads to an additional time-averaging yielding an autocorrelation function of the type 
\begin{equation} 
F(t) = \lim_{T\to\infty}\frac{1}{2T}\int_{-T}^{T} f(t+s)f(s)\,ds\ .
\label{auto}
\end{equation}
Applying this generalization of (\ref{Fer1}) to the Hamiltonian (\ref{hint}) and averaging over the reservoir states as in (\ref{prob}) we obtain the transition probability over sufficiently long time interval
\begin{equation}
{\bar P}_{kl} = \frac{1}{\hbar}
\sum_{\gamma',\gamma}\int d\omega \int dE'\int dE\, \sigma(E)|<E',\gamma'|R|E,\gamma>|^2
G_{kl}(\omega)\delta (E - E' +\hbar\omega)
\nonumber
\end{equation}
\begin{equation}
= \frac{1}{\hbar^2}
\sum_{\gamma',\gamma} \int dE'\int dE\, \sigma(E)|<E',\gamma'|R|E,\gamma>|^2
G_{kl}\bigl((E' - E)/\hbar \bigr)
\label{bb}
\end{equation}
where $G_{kl}(\omega)$ is the power spectrum of the autocorrelation function
\begin{equation} 
F_{kl}(t) = \lim_{T\to\infty}\frac{1}{2T}\int_{-T}^{T} <k|S(t+u)|l><l|S(u)|k>\,du
= \int_{-\infty}^{\infty}G_{kl}(\omega)e^{i\omega t}\,d\omega\ .
\label{aubb}
\end{equation}
One should notice that by introducing a time-dependent Hamiltonian $H_S(t)$ we modify the energy balance. For a constant
$H_S$ only the frequencies $\omega = (\epsilon_k -\epsilon_l)/\hbar$ can appear in the power spectrum $G_{kl}(\omega)$. Now, the power spectrum can be altered by adding or extracting the energy quanta $\hbar\omega'$
supplied by the varying external field present in $H_S(t)$.\\
The main idea of the "bang-bang" technique is to shift the support of the power spectrum $G_{kl}(\omega)$ 
beyond the cut-off value $E_{cut}$ in order to make the energy conservation condition $E'-E=\hbar\omega$ impossible to satisfy with $\omega$ within this support. In other words the overlap of two functions in the final integral (\ref{bb}) must be very small \cite{Ov}.\\ 
In principle it can be done using a Hamiltonian $H_S(t)$ which contains terms
rapidly oscillating with the frequency $\Omega >> E_{cut}/\hbar$. Take a simple example of the Hamiltonian periodic in time $H_S(t+\tau)= H_S(t)$. For the eigenvectors of the {\em Floquet operator}
\begin{equation} 
U(\tau)\equiv U(\tau,0)\ ,\  U(\tau)|k>= \exp \bigl(-\frac{i\epsilon_k \tau}{\hbar}\bigr)|k>
\label{pro}
\end{equation}
where
\begin{equation} 
U(t,s) ={\mathbb T}\exp\Bigl(-\frac{i}{\hbar}\int_s^t H_S(u)du\Bigr)
\label{pr1}
\end{equation}
we have the following representation of the evolution \cite{CFKS}
\begin{equation} 
U(t,0)|k> = \exp \bigl(-\frac{i\epsilon_k t}{\hbar}\bigr)|\phi_k(t)>\ .
\label{pr2}
\end{equation}
Here $\phi_k(t)$ is a periodic (with a period $\tau$) function of $t$ with values in normalized vectors of the Hilbert space. From the periodicity of $\phi_k(t)$ it follows that (see (\ref{aubb}))
\begin{equation} 
F_{kl}(t) = \exp\bigl\{-\frac{i}{\hbar}(\epsilon_l -\epsilon_k) t\bigr\}\times
\nonumber
\end{equation}
\begin{equation} 
\times \frac{1}{\tau}\int_0^{\tau} <\phi_k(t+s)|S|\phi_l(t+s)><\phi_l(s)|S|\phi_k(s)>\,ds\ .
\label{aubb1}
\end{equation}
and the term given by the integral is periodic in $t$.
As a consequence the power spectrum possesses the following structure
\begin{equation} 
G_{kl}(\omega) = \sum_{n=-\infty}^{\infty}\nu_n \,\delta \bigl[\omega -\frac{i}{\hbar}(\epsilon_l -\epsilon_k) - \frac{2\pi n}{\tau}\bigr]
\label{aubb2}
\end{equation}
with weights $\nu_n \geq 0$ depending on the particular choice of $H_S(t)$.
Therefore, decreasing the period $\tau$ below the value $\hbar/E_{cut}$ and designing a proper shape of  $H_S(t)$ we can, in principle, reduce transition probabilities ${\bar P}_{kl}$.
However, one should notice that the similar effect can be achieved, simply, by performing very fast gates with the duration time $t_g << \hbar/E_{cut}$. 

Unfortunately, for most of the physically relevant models the cut-off energy provides a maximal energy scale for which the given model of a system-reservoir interaction is valid. For example, linear coupling to phonons is restricted by the Debye energy $E_{cut}= \hbar\omega_D$, in quantum
optics such a cut-off is provided by the ionisation energy of atoms while in  quantum electrodynamics
one usually takes $E_{cut}= m_ec^2$ , where $m_e$ is the electron mass. To preserve the consistency of the mathematical model we should not consider the frequencies satisfying
$\hbar\omega>> E_{cut}$. If such frequencies can be physically realized then usually different mechanisms of decoherence must be also taken into account (e.g. nonlinear, 2-phonon processes in solid state).

Perhaps, the only example for which the above objections do not apply is the leading decoherence mechanism in NMR systems. It is related to the fluctuations of the local magnetic field possessing a natural cut-off in time-scale and hence energy scale as well. The same holds for the "engineered noises" used is some of the experiments on quantum error control \cite{F}.

\section{Zeno effect}

In the early discussions of the {\em Zeno effect} its physical origin was attributed to the frequent von Neumann projective measurements, described by the projections $P_j$ ($\sum_j P_j = {\bf 1}, P_jP_k=\delta_{jk}P_j$); such projections combined with the Hamiltonian evolution $U(t)= \exp \{-(i/\hbar) Ht\}$ produce an effective dynamics given by
the asymptotic formula
\begin{equation} 
\rho (t) = \sum_j W_j(t)\rho W_j^{\dagger}(t)\ ,
\label{Zeno}
\end{equation}
where
\begin{equation}
W_j(t)=\lim_{n\to\infty}\bigl[P_jU(t/n)P_j\bigr]^n=
P_j \exp\bigl\{-(i/\hbar) P_jHP_jt\bigr\}\ .
\label{Zeno1}
\end{equation}
The main consequence of frequent measurements is the stabilization of a quantum state which is trapped into a subspace
corresponding to the projector $P_j$. This idea has been used, for example, to describe the suppression of decay
of an unstable particle when continuously observed. Some authors went even further to apply Zeno effect in the case of "observation without observer", i.e. for a general open system for which interaction with an environment replaces  continuous measurements. However, the predictive power of such a theory is very weak. Take a neutron, which is unstable as a free particle. Inside a nucleus the interaction with other nucleons can stabilize a neutron or not.
Similarly, a proton, a stable particle by its own, can become unstable when put inside a nucleus. To predict the behavior of a nucleon we need to know the detailed energy balance of the total composed system and the rough picture of  "Zeno effect" for a continuously perturbed ("observed") particle is rather misleading.

In more recent papers the phenomenological picture of wave collapse after von Neumann measurement has been replaced
by more fundamental models of a system strongly interacting with an environment represented either by a large reservoir
(e.g. macroscopic measuring apparatus) or periodic external fields. Our aim is to provide an elementary explanation based on the energy conservation principle.
 
\subsection{Open system in strong coupling regime}

We consider again the model of an open system with the Hamiltonian (\ref{open}) but with the assumption that the interaction with the environment is strong. The Hilbert space basis of a system $\{|j>\}$ is now determined by $S$ and not by $H_S$ 
i.e.
\begin{equation}
S = \sum_j s_j P_j \ , P_j= |j><j|\ .
\label{sdec}
\end{equation}
We use the notation
\begin{equation}
H_R^{(j)} = H_R + s_j R\ , \ H_S = \sum_j \epsilon_j P_j + V   
\label{str}
\end{equation}
where $<j|V|j>=0$. We can decompose the total Hamiltonian as
\begin{equation}
H= H_0 + V \ ,\ \ H_0 = \sum_j P_j\otimes(\epsilon_j {\bf 1}_R + H_R^{(j)} )   
\label{str1}
\end{equation}
where the off-diagonal part of $H_S$, denoted by $V$, is treated as a small perturbation.\\
The eigenvectors and the eigenvalues of $H_R^{(j)}$ are given by
\begin{equation}
H_R^{(j)}|E,\gamma;j> = E |E,\gamma;j> \ .   
\label{eig}
\end{equation}
We can assume that for any $j$ , $E\geq E^{(j)}_{g}$ where $E^{(j)}_{g}$ is the ground state energy
of  $H_R^{(j)}$. As we assume the strong coupling between {\cal S} and {\cal R} the ground state energies $E^{(j)}_g$ depend strongly on the index "j". We can also find the eigenvectors and the eigenvalues of $H_0$ 
\begin{equation}
H_0 |j>\otimes|E,\gamma;j> = (\epsilon_j +E )|j>\otimes|E,\gamma;j>\ ,\ E\geq E^{(j)}_{g}\ .    
\label{eig1}
\end{equation}
Assuming for simplicity the zero temperature reservoir (or in other words initial non-degenerated ground state $|E^{(l)}_{g}>$), we can compute the transition probability $P_{kl}$ averaging over the final states of the environment
\begin{equation}
P_{kl} = \frac{2\pi}{\hbar}|<k|V|l>|^2 
\sum_{\gamma}\int_{E^{(k)}_{g}}^{\infty} dE\, |<E,\gamma|R|E^{(l)}_{g}>|^2\delta(\epsilon_k + E - \epsilon_l - E^{(l)}_{g})\ .
\label{przeno}
\end{equation}
If the following {\em threshold condition} is satisfied
\begin{equation}
E^{(k)}_{g}+\epsilon_k > E^{(l)}_{g}+\epsilon_l
\label{thr}
\end{equation}
then the energy conservation forbids
the transition $|l> \mapsto |k>$, i.e.  $P_{kl} = 0$ \cite{Z}.

\subsection{A model of Zeno effect}

We discuss now a simplified model describing the experimental test of the Zeno effect \cite{I}. A 3-level atom is subjected
to the interaction with a strong laser field driving the optical transition between the levels 1, 3 and to a radiofrequency
field causing  Rabi oscillations with a frequency $\Omega$ between the levels 1, 2. The model Hamiltonian can be written as
\begin{equation}
H = \hbar\omega_{13}|3><3| +  \frac{\hbar}{2}\Omega(|1><2| + |2><1|) 
+ \hbar\sum_k \omega_k \bigl[a^{\dagger}- {\bar F}_k\bigr]\bigl[a_k - F_k\bigr]
\nonumber
\end{equation}
\begin{equation}
+ g_{13}(|1><3|+ |3><1|)\otimes \sum_k (f_k a_k + {\bar f}_k a_k^{\dagger})\ .
\label{modzen}
\end{equation}
Here $a_k^{\dagger}, a_k$ are creation and annihilation operators of the electromagnetic field and $F_k$ correspond to a {\em classical field} describing the strong pumping mechanism of the laser. This classical field  drives the quantum electromagnetic field into a new ground state - a coherent state $|F>$ satisfying $a_k |F> = F_k |F>$. It is convenient to introduce a new atomic basis 
$ |\pm> = \frac{1}{\sqrt 2}(|1> \pm |3>), |2>$ and new field operators  $b_k = a_k-F_k$.

The Hamiltonian can be now written in a form (\ref{str1}) 
\begin{equation}
H= H_0 + V\ ,\ H_0 = P_-\otimes(H_-^{(em)} + \epsilon_-)+ P_+\otimes(H_+^{(em)} + \epsilon_+) + P_2\otimes H^{(em)}  
\label{str2}
\end{equation}
where $P_{\pm}= |\pm><\pm|, P_2=|2><2|$ and
\begin{equation}
H^{(em)}= \sum_k \omega_k b^{\dagger}_k b_k \ ,\ H_{\pm}^{(em)}= H^{(em)}\pm g_{13} \sum_k (f_k b_k + {\bar f}_k b_k^{\dagger})\ ,   
\label{str3}
\end{equation}
\begin{equation}
\epsilon_{\pm}= \frac{\hbar\omega_{13}}{2}\pm g_{13} \sum_k (f_k F_k + {\bar f}_k {\bar F}_k)\ ,   
\label{str4}
\end{equation}
\begin{equation}
V= -\frac{\hbar\omega_{13}}{2}(|+><-| + |-><+|) 
\nonumber
\end{equation}
\begin{equation}
+\frac{\hbar\Omega}{2\sqrt{2}}(|+><2| + |-><2|+ |2><-| + |2><+|)\ . 
\label{str5}
\end{equation}
The ground state energies of the Hamiltonians $H^{(em)}, H_{\pm}^{(em)}$ are given by
\begin{equation}
E_g = 0 \ ,\ E_g^{(-)} = E_g^{(+)}= -\sum_k\frac{|f_k|^2}{\hbar^2\omega_k^2}
\label{gre}
\end{equation}
respectively with the corresponding coherent ground states $|F> ,|F^{(\pm)}>$ satisfying
\begin{equation}
b_k|F> = 0\ , b_k |F^{(\pm)}>= \mp \frac{\bar f_k}{\hbar\omega_k}|F^{(\pm)}>\ .
\label{grst}
\end{equation}
Assume that $g_{13} \sum_k (f_k F_k + {\bar f}_k {\bar F}_k) >0$. Then, the energy of the ground state $|->\otimes |F^{(-)}>$ of $H_0$, for a strong laser field $\{F_k\}$, is much lower than the energy of the $H_0$- eigenstate $|2>\otimes |F>$. Therefore, according to (\ref{przeno},\ref{thr}) the transition from the state $|->$ to the state $|2>$ driven by the Rabi term  $\frac{\hbar}{2}\Omega(|1><2| + |2><1|)$ is forbidden by the energy conservation principle.\\
This phenomenon is interpreted as the Zeno effect caused by continuous measurements performed on the levels 1 and 3 by means of the interaction with a
strong electromagnetic (laser) field.

The main  difficulty in application of the Zeno effect in quantum information processing is the presence
of a strong coupling with environment. The structure of the unperturbed Hamiltonian (\ref{str1},\ref{eig1}) suggests that a subsystem {\cal S} becomes a {\em dressed system} and the very definition of 
the collection of "individual qubits" with its own controlled unitary dynamics becomes questionable.

\section{Decoherence free subspaces and subsystems}

Consider a model of an open system {\cal S} weakly interacting with a reservoir {\cal R} as in Section 2.1.
The physical mechanism leading to decoherence-free subspaces (DFS) is entirely due to the vanishing matrix
elements $<k|S|l>=0$ for all pairs of  states, such that $|l>$  belongs to a certain subspace ${\cal H}^{DFS}$ of the Hilbert space of {\cal S} and $|k>$ is an arbitrary state orthogonal to $|l>$. This happens if and only if  $S|l> = s|l>$ for all $|l>\in {\cal H}^{DFS}$, i.e. the operator $S$ acts as a scalar on the subspace ${\cal H}^{DFS}$. In such a case we can say that the transitions  $|l> \leftrightarrow |k>$ are {\em forbidden}.

\subsection{Symmetries and forbidden transitions}

Typically, decoherence free subspaces are associated with the symmetries characterizing the system and its interaction
with the environment \cite{RA,ZF,LW}. Assume that these symmetries are given in terms of a generally reducible representation of a certain semi-simple Lie group acting on the Hilbert space of the system {\cal S}. This representation is given by a basis of operators $X_{\mu}=X_{\mu}^{\dagger}$ generating
the representation of the corresponding Lie algebra and choosen in such a way that the Casimir operator $C$ can be written as \cite{H}
\begin{equation}
C = \sum_{\mu}X_{\mu}^2\ .
\label{cas}
\end{equation}
Denote by ${\cal H}_0$ the subspace spanned by the eigenvectors of $C$ with the eigenvalue zero. It follows that for any
$\phi\in{\cal H}_0$ we have
\begin{equation}
0=<\phi |C|\phi>= \sum_{\mu}\|X_{\mu}\phi\|^2 \Leftrightarrow X_{\mu}\phi =0\ .
\label{cas1}
\end{equation}
Therefore for any interaction Hamiltonian (\ref{open}) which is of the form
\begin{equation}
H_{int} =\sum_{\mu} X_{\mu}\otimes R_{\mu}
\label{cas2}
\end{equation}
the subspace ${\cal H}_0$ is decoherence-free.\\
Typical examples are provided by systems with rotational (or isospin) symmetry. They are equipped with a (generally reducible)
representation of the $SU(2)$ group and corresponding "angular momentum" operators $[J_k, J_l] = i\epsilon_{klm}J_m$, $k,l,m= 1,2,3,$
and the Casimir operator ${\bf J}^2 = J_1^2 + J_2^2 +J_3^2$ with the eigenvalues $j(j+1)$, $j=0,1/2,1,...$. In this case all {\em singlet states} corresponding to $j=0$ span the DFS for the interactions with an environment of the form 
$\sum_k J_k\otimes R_k$.

\subsection{Collective decoherence}

The most popular model leading to DFS consists of $2N$ qubits coupled to environment by means of a
{\em collective interaction} ( for the related superradiance and subradiance phenomena, see \cite{GH, RA1})
\begin{equation}
H_{int} =\sum_{j=1}^{2N}\sum_{k=1}^3 \sigma_k^{(j)}\otimes R_k 
\label{col}
\end{equation}
where $\sigma_k^{(j)},$ are  Pauli matrices for the $j$-th qubit. The operators
$J_k = \frac{1}{2}\sum_{j=1}^{2N}\sigma_k^{(j)}$, $k=1,2,3$  generate a reducible representation of $SU(2)$ an therefore
the $2N$-qubit singlet states span DFS of  dimension $(2N!)/(N+1)!N!$. Here the mechanism leading to DSF is due to the symmetry of the interaction Hamiltonian with respect to qubits permutations.

\subsection{Decoherence-free subsystems}

A generalization of  DFS leads to the notion of {\em decoherence-free subsystems}\cite{LW}. One assumes that the operator
$S$ can be represented in the following block-diagonal form
\begin{equation}
S = \bigoplus_{J} {\bf I}_{n_J}\otimes S_J\ .
\label{NS}
\end{equation}
corresponding to the decomposition of the Hilbert space
\begin{equation}
{\cal H}_S = \bigoplus_{J} {\bf C}^{n_J}\otimes{\bf C}^{d_J}
\label{NSH}
\end{equation}
typically associated with reducible representations of compact Lie groups.
It means that $S$ acts as a scalar on Hilbert spaces ${\bf C}^{n_J}$ which now correspond to certain fictitious  subsystems. According to the previous discussion these subsystems are decoherence-free.

As argued in \cite{ZL} the idea of topologically protected states \cite{Kit} is also closely related to decoherence-free subsystems. 
\subsection{Physical implementation}
The engineering of the interaction Hamiltonian (\ref{col}) in  real systems is a very difficult task. Namely, the permutation invariance can be only approximative and the operators $R_k$ always depend on $j$ also. To reduce this dependence one should place all qubits within 
a space region of the linear dimension which is small in comparison with the typical wavelength of the reservoirs fluctuations coupled to the qubit system. On the other hand, a dense packing of qubits leads to unwanted interactions between them which are not permutation invariant and hence spoil the effect of collective decoherence suppression.

\section{Conclusions}

We have shown using the elementary quantum mechanical time-dependent perturbation theory that the different methods of controlling decoherence are based on the combination of the energy conservation principle
with the concept of forbidden transitions usually related to symmetries of the system. All those methods possess their own rather restricted ranges of applicability and demand often contradictory means to be implemented. For instance, the Zeno effect is based on a strong interaction with an environment while the other techniques rely on the weak coupling
assumption. Minimal decoherence technique demands coupling to low frequencies of the reservoir, similarly to DFS based
on permutation invariance guaranteed by the coupling to long wave fluctuations. On the other hand "bang-bang" techniques
employ high frequency regime. One can expect that only an optimized combination of different techniques might be, to some extent, successful in noise reduction for microscopic quantum devices (see \cite{DA} for the overview of {\em hybrid methods}).

The author is grateful to Daniel Lidar and  Marco Piani for very useful comments. This work was supported by the Polish Ministry of Scientific Research under Grant 2P03B 084 25 
and by EC grant RESQ IST-2001-37559.


\begin{thebibliography}{9}

\bibitem {NC}
M.A. Nielsen and I.L. Chuang: \emph{Quantum Computation and Quantum Information}, 
(Cambridge University Press, Cambridge, 2000).

\bibitem {A}
G. Alber et.al.:\emph{Quantum Information} (Springer, Berlin 2001)

\bibitem{BF}
F. Benatti and R. Floreanini (Eds.):\emph{Irreversible Quantum Dynamics}, LNP 622,
(Springer, Berlin, 2003)

\bibitem{J}
E. Joos et.al.:\emph{Decoherence and the Appearance of the Classical World in Quantum Theory} (Second Edition),
(Springer, Berlin 2003)

\bibitem{RA}
R. Alicki: article in \cite{BF} and references therein.

\bibitem {AHHH}
R. Alicki, M. Horodecki, P. Horodecki and R. Horodecki: Phys.Rev. {\textbf A 65}, 062101 (2002)


\bibitem{VL}
L. Viola and S. Lloyd: Phys.Rev.\textbf{A 58}, 2733 (1998)

\bibitem {MS}
B. Misra and E.C.G. Sudarshan.: J. Math. Phys \textbf{18}, 756 (1977)

\bibitem {FP}
P. Facchi and S. Pascazio, article in \cite{BF} an references therein.

\bibitem {ZF}
P. Zanardi: Phys.Rev.\textbf{A 63}, 012301-1 (2001);
S. De Filippo: Phys.Rev.\textbf{A 62}, 052307-1 (2000)

\bibitem{LW}
D.A. Lidar and K.B. Whaley: article in \cite{BF} and references therein.

\bibitem{K}
E. Knill, et.al: \emph{Introduction to Quantum Error Correction}, in Quantum Computation, ed. S.J. Lamonaco  Jr., pp.221-235 (AMS, Providence 2002), arXiv:quant-ph/0207170 , (2002)

\bibitem{ZLF}
P. Zanardi and S. Lloyd: Phys.Rev.Lett.\textbf{90}, 067902 (2003);
P. Facchi, D.A. Lidar and S. Pascazio:Phys.Rev.\textbf{A 69}, 032314 (2004)

\bibitem{F}
E.M. Fortunato et.al.: New J. Phys.\textbf{4}, 5 (2002)

\bibitem{KTH}
R. Kubo et.al.: \emph{Statistical Physics II}, (Springer, Berlin, 1985)

\bibitem{Ov}
The similar ideas were discussed in \cite{AHHH} and using different techniques
by Kofman and Kurizki: Phys.Rev.Lett.\textbf{93}, 130406 (2004)
\bibitem{CFKS}
H.L. Cycon et.al.: \emph{Schr\"odinger Operators}, p.147,
(Springer, Berlin, 1987)

\bibitem{Z}
In \cite{J} (p.126-133) a similar Hamiltonian model of the Zeno effect is also discussed.
The main advantage of our approach is the detailed analysis of the Hamiltonian spectrum
(\ref{eig1}) leading to the sharp threshold condition (\ref{thr}) while in \cite{J} only a mild suppression of decoherence inversely proportional to the system-environment coupling constant is predicted.

\bibitem{I}
W.M. Itano et.al.: Phys.Rev,\textbf{A 41}, 2295 (1990)

\bibitem{H}
M. Hamermesh: \emph{Group Theory and Its Applications to Physical Problems},
(Addison-Wesley, London, 1964)

\bibitem{GH}
M. Gross and S. Haroche: Phys.Rep.\textbf{93}, 301 (1982)

\bibitem{RA1}
R. Alicki: Physica \textbf{A 150}, 455 (1988)

\bibitem{ZL}
P. Zanardi and S. Lloyd: Phys.Rev.Lett.\textbf{90}, 067902 (2003)

\bibitem{Kit}
A. Yu. Kitaev: Ann. Phys. \textbf{303}, 2 (2003)

\bibitem{DA}
M.S. Byrd and D.A. Lidar: J. Mod. Optics \textbf{50}, 1285 (2003)

\end{thebibliography}
\end{document}